\documentclass[pra, aps, twocolumn, preprintnumbers,showpacs, floatfix,superscriptaddress]{revtex4}

\usepackage{latexsym}
\usepackage{mathrsfs}
\usepackage{amsmath}
\usepackage{amssymb}
\usepackage{amsfonts}
\usepackage{ifthen}

\usepackage{graphics,epsfig}

\def\complex{\mathbb{C}}
\newcommand{\half}{\frac{1}{2}}
\def\integers{\mathbb{Z}}
\newcommand{\bestar}{\begin{equation*}}
\newcommand{\eestar}{\end{equation*}}
\newcommand{\beastar}{\begin{eqnarray*}}
\newcommand{\eeastar}{\end{eqnarray*}}
\newcommand{\cH}{{\cal H}}

\newcounter{protoCount}
\newcounter{protoList}
\newsavebox{\tmpbox}
\newlength{\protobox}
\newenvironment{protocol}[2]{
\bigskip
\addtocounter{protoCount}{1} \noindent \begin{lrbox}{\tmpbox}
\setlength{\protobox}{\columnwidth} \addtolength{\protobox}{-0.5cm}
\begin{minipage}[c]{\protobox}
\begin{bfseries}Protocol \theprotoCount: #1\end{bfseries}
\ifthenelse{\equal{#2}{\empty}}{}{\\Prerequisite: #2}
\begin{list}{\begin{bfseries}\arabic{protoList}:\end{bfseries}}
{\usecounter{protoList}} }{
\end{list}
\end{minipage}\end{lrbox}
\fbox{\usebox{\tmpbox}}
}

\newcommand{\footnoteremember}[2]{
  \footnote{#2}
  \newcounter{#1}
  \setcounter{#1}{\value{footnote}}
}
\newcommand{\footnoterecall}[1]{
  \footnotemark[\value{#1}]
}

\newcommand{\proj}[1]{\ket{#1}\bra{#1}}

\newcommand{\be}{\begin{equation}}
\newcommand{\ee}{\end{equation}}
\newcommand{\bea}{\begin{eqnarray}}
\newcommand{\eea}{\end{eqnarray}}

\newcommand{\hilbert}{\mathcal{H}}
\newcommand{\cY}{{\cal Y}}      
\newcommand{\cU}{{\cal U}}      

\newcommand{\cE}{{\cal E}}

\newcommand{\I}{I_{acc}}        
\newcommand{\ens}{{\cal E}}
\newcommand{\nc}{\newcommand}
\nc{\RR}{{{\mathbb R}}}
\nc{\CC}{{{\mathbb C}}}
\nc{\FF}{{{\mathbb F}}}
\nc{\NN}{{{\mathbb N}}}
\nc{\ZZ}{{{\mathbb Z}}}
\nc{\PP}{{{\mathbb P}}}
\nc{\QQ}{{{\mathbb Q}}}
\nc{\UU}{{{\mathbb U}}}
\nc{\EE}{{{\mathbb E}}}
\nc{\id}{{\mathbb I}}
\nc{\polylog}{\operatorname{polylog}}
\def\e{\epsilon}
\nc{\rnc}{\renewcommand} \nc{\beq}{\begin{equation}}
\nc{\eeq}{{\end{equation}}} \nc{\beqa}{\begin{eqnarray}}
\nc{\eeqa}{\end{eqnarray}} \nc{\lbar}[1]{\overline{#1}}
\nc{\avg}[1]{\langle#1\rangle} \rnc{\max}{\operatorname{max}}
\nc{\rank}{\operatorname{rank}}
\nc{\smfrac}[2]{\mbox{$\frac{#1}{#2}$}} \nc{\ox}{\otimes}
\nc{\dg}{\dagger} \nc{\dn}{\downarrow} \nc{\cA}{{\cal A}}
\nc{\optimal}{^*} \nc{\csupp}{{\operatorname{csupp}}}
\nc{\qsupp}{{\operatorname{qsupp}}}
\nc{\esupp}{{\operatorname{esupp}}} \nc{\var}{\operatorname{var}}
\nc{\rar}{\rightarrow} \nc{\lrar}{\longrightarrow}
\nc{\poly}{\operatorname{poly}}

\usepackage{amsfonts}

\def\01{\{0,1\}}

\newcommand{\eps}{\varepsilon}
\newcommand{\ket}[1]{|#1\rangle}
\newcommand{\bra}[1]{\langle#1|}
\newcommand{\outp}[2]{|#1\rangle\langle#2|}

\newcommand{\braket}[2]{\langle #1|#2\rangle}

\newcommand{\Tr}{\mbox{\rm Tr}}

\newtheorem{definition}{Definition}
\newtheorem{theorem}{Theorem}
\newtheorem{lemma}{Lemma}
\newtheorem{proposition}{Proposition}
\newtheorem{corollary}{Corollary}

\newenvironment{proof}
{\noindent {\bf Proof. }}
{{\hfill $\Box$}\\
 \smallskip}

\begin{document}

\title{Possibility, Impossibility and Cheat-Sensitivity of Quantum Bit String Commitment}

\author{Harry \surname{Buhrman}}
\affiliation{CWI, Kruislaan 413, 1098 SJ
Amsterdam, The Netherlands}

\author{Matthias \surname{Christandl}}
\email[]{matthias.christandl@qubit.org}
\affiliation{DAMTP, University of Cambridge, Wilberforce Road, Cambridge, CB3
0WA, U.K.}

\author{Patrick \surname{Hayden}}
\affiliation{School of Computer
Science, McGill University, Montreal, Canada}

\author{Hoi-Kwong \surname{Lo}}
\affiliation{Department of
ECE and Physics,
University of Toronto, Canada M5G 3G4}

\author{Stephanie \surname{Wehner}}
\email[]{wehner@cwi.nl}
\affiliation{CWI, Kruislaan 413, 1098 SJ
Amsterdam, The Netherlands}

\date{\today}

\begin{abstract}
Unconditionally secure non-relativistic bit commitment is known to
be impossible in both the classical and the quantum worlds. But
when committing to a string of $n$ bits at once, how far can we
stretch the quantum limits? In this paper, we introduce a framework
for quantum schemes where Alice commits a string of $n$ bits to Bob
in such a way that she can only cheat on $a$ bits and Bob can learn
at most $b$ bits of information before the reveal phase. Our results
are two-fold: we show by an explicit construction that in the
traditional approach, where the reveal and guess probabilities form
the security criteria, no good schemes can exist: $a+b$ is at least
$n$. If, however, we use a more liberal criterion of security, the
accessible information, we construct schemes where $a=4\log_2
n+O(1)$ and $b=4$, which is impossible classically. We furthermore present
a cheat-sensitive quantum bit string commitment protocol for which we
give an explicit tradeoff between Bob's ability to gain information
about the committed string, and the probability of him being detected cheating.
\end{abstract}

\maketitle

Commitments play an important role in modern day cryptography.
Informally, a commitment allows one party to prove
that she has made up her mind and cannot
change it, while hiding the actual decision until later.
Imagine two mutually distrustful parties Alice and Bob at distant
locations. They can only communicate over a channel, but want to
play the following game: Alice secretly chooses a bit $x$. Bob wants
to be sure that Alice indeed has made her choice. At the same time, Alice wants
to keep $x$ hidden from Bob until she decides to reveal $x$. To
convince Bob that she made up her mind, Alice sends Bob a
commitment. From the commitment alone, Bob cannot deduce $x$. At a
later time, Alice reveals $x$ and enables Bob to open the
commitment. Bob can now check if Alice is telling the truth. This
scenario is known as \emph{bit commitment}.

Bit commitment is a very powerful cryptographic primitive with a
wide range of applications. It has been shown that quantum oblivious
transfer (QOT)~\cite{crepeau:practicalOT} can be achieved provided
there exists a secure bit commitment
scheme~\cite{yao:otFromBc,crepeau:qot}. In turn, oblivious transfer
is known to be sufficient for solving the general problem of secure
two-party
computation~\cite{crepeau:committedOT,kilian:foundingOnOT}.
Commitments are also useful for constructing zero-knowledge
proofs~\cite{goldreich:book1}. Furthermore, a bit commitment
protocol can be used to implement secure coin
tossing~\cite{blum:coin}. Classically, unconditionally secure bit
commitment is known to be impossible. Unfortunately after several
quantum schemes were
suggested~\cite{bb84,brassard:bcAndCoin,brassard:bc},
non-relativistic quantum bit commitment has also been shown to be
impossible
~\cite{mayers:trouble,lo&chau:bitcom,mayers:bitcom,lo:coin,
brassard:brief,lo:promise}. Only very limited degrees of concealment
and binding can be achieved~\cite{spekkens:tradeoffBc}. In the face
of these negative statements, what can we still hope to achieve?

\subsection{String Commitment}
Here we take a different approach
and look at the task of committing to a string of $n$ bits at once in the setting
where Alice and Bob have unbounded resources. Since perfect bit commitment is impossible, perfect
string commitment is impossible, too. However, is it
possible to design meaningful string commitment schemes when we
allow for a small ability to cheat on both Alice's and Bob's side?
To make this question precise, we introduce a framework for the
classification of string commitments in terms of the length $n$ of the string,
Alice's ability to cheat on $a$ bits and Bob's ability to acquire
$b$ bits of information before the reveal phase.
Instead of asking for a perfectly binding commitment, we allow Alice
to reveal up to $2^a$ strings successfully: Bob will accept any such
string as a valid opening of the commitment.
Formally, we demand that $\sum_{x \in \01^n}
p^A_x \leq 2^a$, where $p^A_x$ is the probability that
Alice successfully reveals string $x$ during the reveal phase.
Contrary to classical computing, Alice can always choose
to perform a superposition of string commitments without Bob's knowledge.
Thus even for a perfectly
binding string commitment we would only demand $\sum_{x \in
\01^n}p^A_x \leq 1$, since a strategy based on
superpositions is indistinguishable from the ``classical''
honest behaviour of choosing a string beforehand and then
committing to it. At the same time, we relax Bob's security
condition, and allow him to acquire at most $b$ bits of
information before the reveal phase.
The nature of his
security definition is crucial to our investigation: If $b$
determines a bound on his probability to guess Alice's string, then
we prove that $a + b$ is at least $n$ (up to a small constant).
We write $(n,a,b)$-QBSC for a quantum bit string commitment protocol
where the string has length $n$ and $a$ and $b$ are the security
parameters for Alice and Bob as explained in detail below.
In Section~\ref{imposs}, we show\\

\begin{flushleft}
{\bf Impossibility of $(n,a,b)$-QBSC:}\\
Every $(n,a,b)$-QBSC scheme with $a+b + c < n$ is insecure, where
$c \approx 7.61$.
\end{flushleft}

Our proof makes use of privacy amplification with two-universal hash functions.
If the protocol is executed multiple times in parallel, we prove that any quantum bit string commitment protocol with $a + b <n$ is insecure.
We refer to these results as ``impossibilities'',
as they show that QBSCs offer almost no
advantage over the trivial classical protocol:
Alice first sends
$b$ bits of the $n$ bit string to Bob during the commit phase, and
then supplies him with the remaining $n-b$ bits in the reveal
phase.

The second part of the paper is devoted to the ``possibility'' of
QBSC.
If we weaken our standard of security
and measure Bob's information gain in terms of
the accessible information, it becomes possible to construct
meaningful QBSC protocols with $a = 4 \log_2 n + O(1)$ and $b = 4$. Our protocols are based on the
effect of locking classical information in
quantum states~\cite{terhal:locking}. This surprising effect shows that
given an initial shared quantum state, the transmission of $\ell$ classical
bits can increase the total amount of correlation by more than $\ell$ bits.
In Section~\ref{protocol}, we show\\

\begin{flushleft}
{\bf Possibility of $(n,a,b)-\mbox{QBSC}_{\I}$:}\\
For $n \geq 3$, there exist $(n,4 \log_2 n + O(1),4)-\mbox{QBSC}_{\I}$
protocols.
\end{flushleft}

We then consider cheat-sensitive protocols: Even though Bob is in principle
able to gain a large amount of information on Alice's committed string,
honest Alice has a decent probability of detecting such an attempt to cheat
the protocol. We give an explicit tradeoff between Bob's information
gain, and Alice's ability to catch him cheating.
In Section~\ref{CSprotocol},
we show\\

\begin{flushleft}
{\bf Possibility of cheat-sensitive $(n,1,n/2)-\mbox{QBSC}_{\I}$:}\\
There exist a $(n,1,n/2)-\mbox{QBSC}_{\I}$ that is cheat-sensitive against Bob.
If Bob is detected cheating with probability less than $\eps$, then his classical information gain is less than
$4 \sqrt{\eps} \log_2 d + 2 \mu(2\sqrt{\eps})$ with $\mu(x) = \min\{-x \log_2 x,1/e\}$.
\end{flushleft}

\subsection{Related Work}
To obtain bit commitment, different restrictions
have been introduced into the model. Salvail~\cite{salvail:physical}
showed that, for any fixed $n$, secure
bit commitment is possible provided that the sender is not able to
perform generalized measurements
on more then $n$ qubits coherently. Large $n$ coherent measurements
are not yet feasible,
so his result provides an implementation which is
secure under a plausible technological
assumption. DiVincenzo, Smolin and Terhal took a different
approach~\cite{terhal:superselectionBc},
showing that if the bit commitment is forced to be
ancilla-free, a type of asymptotic security is still
possible. Bit commitment is also possible if the adversary's
quantum storage is bounded~\cite{serge:bounded,serge:new,js:compose}
or noisy~\cite{noisyOT}.
Classically,
introducing restrictions can also open new possibilities.
Cachin, Cr\'epeau and Marcil have shown
how to implement bit commitment
via oblivious transfer under the assumption
that the size of the receiver's memory is bounded~\cite{cachin:boundedOT}.
Furthermore, the assumption of a noisy channel can be
sufficient for
oblivious transfer~\cite{crepeau:weakenedOT,winter:bcCapacity}.
A new cryptographic task---called cheat-sensitive bit commitment---has been
studied by Hardy and Kent~\cite{kent:cheatSensitive},
as well as Aharanov, Ta-Shma, Vazirani and Yao~\cite{yao:bitEscrow}:
no restrictions are placed on the adversary initially,
but an honest party should stand a good chance
of catching a cheater.
Kent also showed that
bit commitment can be achieved using
relativistic constraints~\cite{kent:relativisticBc}.

Classically, string commitment is directly linked to
bit commitment and no interesting protocols are possible.
Kent~\cite{kent:sc} first asked what kind of quantum
string commitment (QBSC) can be achieved. He gave a protocol under
the restrictive assumption that Alice does not commit to a
superposition~\cite{kent:personal}. His protocol was modified for experimental purposes
by Tsurumaru~\cite{tsu:sc}.

\section{Preliminaries}

\subsection{Framework}
We first formalize the notion of quantum string commitments in a quantum setting.
\begin{definition}\label{securityDef}
\emph{An $(n,a,b)$-\emph{Quantum Bit String Commitment (QBSC)} is a
quantum communication protocol between two parties, Alice (the
committer) and Bob (the receiver), which consists of two phases and
two security requirements.
\begin{itemize}
\item (Commit Phase) Assume that both parties are honest.
Alice chooses a string $x \in \01^n$ with probability $p_x$.
Alice and Bob communicate and at the end Bob holds state $\rho_x$.
\item (Reveal Phase)
If both parties are honest, Alice and Bob communicate and at the end
Bob learns $x$. Bob accepts.
\item (Concealing)
If Alice is honest, $\sum_{x \in \01^n}p^B_{x|x} \leq 2^{b}$, where
$p^B_{x|x}$ is the probability that Bob correctly guesses $x$ before
the reveal phase.
\item (Binding)
If Bob is honest, then for all commitments of Alice: $\sum_{x \in
\01^n} p^A_x \leq 2^a$, where $p^A_x$ is the probability that Alice
successfully reveals $x$.
\end{itemize}}
\end{definition}
We say that Alice \emph{successfully reveals} a string $x$ if Bob accepts
the opening of $x$, i.e. he performs a test depending on the individual
protocol to check Alice's honesty and concludes that she was indeed honest.
Note that quantumly, Alice can always
commit to a superposition of different strings without being
detected. Thus even for a perfectly binding bit string commitment
(i.e. $a=0$) we only demand that $\sum_{x \in \{0,1\}^n}p^A_x \leq 1$,
whereas classically one wants that $p^A_{x'}=\delta_{x, x'}$.
Note that our concealing definition reflects Bob's a priori
knowledge about $x$. We choose an a priori uniform distribution
(i.e. $p_x = 2^{-n}$) for $(n,a,b)$-QBSCs, which naturally comes from the
fact that we consider $n$-bit strings. A generalization to any
$(P_X,a,b)$-QBSC where $P_X$ is an arbitrary distribution is
possible but omitted in order not to obscure our main line of
argument.
Instead of Bob's guessing probability, one can take any information
measure $B$ to express the security against Bob. In general, we consider an
$(n,a,b)$-$\mbox{QBSC}_B$ where the new concealing condition
$B(\ens) \leq b$ holds for any ensemble $\ens = \{p_x,\rho_x\}$ that Bob can obtain
by a cheating strategy. In the
latter part of this paper we show that for $B$ being the
\emph{accessible information} non-trivial protocols, i.e. protocols
with $a+b \ll n$, exist. The accessible information is defined as
$I_{acc}(\ens) = \max_{M} I(X;Y)$, where $P_X$ is the prior
distribution of the random variable $X$, $Y$ is the random variable
of the outcome of Bob's measurement on $\ens$, and the
maximization is taken over all measurements $M$.

\subsection{Model}
We work in the model of two-party non-relativistic quantum protocols
of Yao~\cite{yao:otFromBc} and then simplified by Lo and
Chau~\cite{lo&chau:bitcom} which is usually adopted in this context.
Here, any two-party quantum protocol can be regarded as a pair of
quantum machines (Alice and Bob), interacting through a quantum
channel. Consider the product of three Hilbert spaces $\hilbert_A$,
$\hilbert_B$ and $\hilbert_C$ of bounded dimensions representing the
Hilbert spaces of Alice's and Bob's machines and the channel,
respectively. Without loss of generality, we assume that each
machine is initially in a specified pure state. Alice and Bob
perform a number of rounds of communication over the channel. Each
such round can be modeled as a unitary transformation on
$\mathcal{H}_A \otimes \mathcal{H}_C$ and $\mathcal{H}_B \otimes
\mathcal{H}_C$ respectively. Since the protocol is known to both
Alice and Bob, they know the set of possible unitary transformations
used in the protocol. We assume that Alice and Bob are in possession
of both a quantum computer and a quantum storage device. This
enables them to add ancillae to the quantum machine and use
reversible unitary operations to replace measurements. By doing so,
Alice and Bob can delay measurements and thus we can limit ourselves
to protocols where both parties only measure at the very end.
Moreover, any classical computation or communication that may occur
can be simulated by a quantum computer. Furthermore, any
probabilistic operation can be modeled as an operation that is
conditional on the outcome of a coin flip. Instead of a classical
coin, we can use a quantum coin and in this way keep the whole
system fully quantum mechanical.

\subsection{Tools}
We now gather the essential ingredients for our proof.
First, we show that every $(n,a,b)$-$\mbox{QBSC}$ is an
$(n,a,b)$-$\mbox{QBSC}_\xi$. The security measure $\xi(\ens)$ is
defined by
\begin{equation} \label{xi}
\xi (\ens)\equiv n- H_2(\rho_{AB}|\rho),
\end{equation}
where $\rho_{AB} = \sum_x p_x \outp{x}{x} \otimes \rho_x$ and $\rho
= \sum_x p_x \rho_x$ are only dependent on the ensemble $\ens=\{p_x,
\rho_x\}$. $H_2(\cdot|\cdot)$ is an entropic quantity defined
in~\cite{renato:diss} $H_2(\rho^{AB}|\rho)\equiv -\log\Tr ((\id
\otimes \rho^{-\frac{1}{2}})\rho_{AB})^2.$ This quantity is directly
connected to Bob's maximal average probability of successful guessing
the string:
\begin{lemma} \label{guessing-lemma}
Bob's maximal average probability of successfully guessing
the committed string,~i.e. $\sup_M \sum_x p_x p^{B, M}_{x|x}$ where $M$
ranges over all measurements and $p^{B, M}_{y|x}$ is the conditional
probability of guessing $y$ given $\rho_x$, obeys
$$
\sup_M \sum_x p_x p^{B, M}_{x|x} \geq 2^{- H_2(\rho_{AB}|\rho)}.
$$
\end{lemma}
\begin{proof}
By definition the maximum average guessing
probability is lower bounded by the average guessing probability for
a particular measurement strategy. We choose the \emph{square-root
measurement} which has operators $M_x = p_x \rho^{-\frac{1}{2}}
\rho_x \rho^{-\frac{1}{2}}$. $p^B_{x|x} = \Tr(M_x \rho_x)$ is the
probability that Bob guesses $x$ given $\rho_x$, hence
\begin{eqnarray*}
\log_2\sum_x p_x p^{B,\max}_{x|x}
&\geq& \log_2\sum_x p_x^2 \Tr(\rho^{-\frac{1}{2}} \rho_x \rho^{-\frac{1}{2}} \rho_x)\\
&=& \log\Tr \left(\left[(\id \otimes \rho^{-\frac{1}{2}})\rho_{AB}\right]^2\right)\\
&=& - H_2(\rho_{AB}|\rho)
\end{eqnarray*}
\end{proof}
Related estimates were derived in~\cite{barnumknill}. For the
uniform distribution $p_x = 2^{-n}$ we have from the concealing
condition that $\sum_x p^B_{x|x} \leq 2^{b}$ which by
Lemma~\ref{guessing-lemma} implies $\xi(\ens) \leq b$ and hence the
following lemma.

\begin{lemma}
Every $(n,a,b)$-QBSC is an $(n,a,b)$-$\mbox{QBSC}_\xi$.
\end{lemma}

Furthermore, we make use of the following theorem, known as \emph{privacy
amplification against a quantum adversary}. In our case, Bob holds
the quantum memory and privacy amplification is used to find Alice's
attack.
\begin{theorem}[Th.~5.5.1 in~\cite{renato:diss} (see
also~\cite{KoMaRe05})]\label{theorem:renato} Let $\mathcal{G}$ be a
class of two-universal hash functions
from $\01^n$ to $\01^s$. Application of $g \in
\mathcal{G}$ to the random variable $X$ maps the ensemble
$\ens=\{p_x, \rho_x\}$ to $\cE_g=\{q^g_y, \sigma^g_y\}$ with
probabilities $q^g_y = \sum_{x \in g^{-1}(y)} p_x$ and quantum
states $\sigma^g_y = \sum_{x \in g^{-1}(y)} p_x \rho_x$. Then \be
\label{eq-renner-koenig} \frac{1}{|\mathcal{G}|} \sum_{g \in
\mathcal{G}} d(\cE_g)  \leq
\frac{1}{2}2^{-\frac{1}{2}[H_2(\rho_{AB}|\rho) - s]},\ee where
$d(\cE) \equiv \delta\big(\sum_x p_x \proj{x} \otimes \rho_x,
\id/2^n\otimes \rho\big)$ (and similarly for $d(\cE_g)$) and
$\delta(\alpha, \beta)\equiv\frac{1}{2} ||\alpha-\beta||_1$ with
$||A||_1 = \Tr\sqrt{A^\dagger A}$.
\end{theorem}

Finally, the following reasoning, previously used to prove the impossibility of
quantum bit commitment~\cite{lo&chau:bitcom, mayers:trouble}, will be
essential:
Suppose $\rho_0$ and $\rho_1$ are density operators that correspond
to a commitment of a ``0'' or a ``1'' respectively. Let
$\ket{\phi_0}$ and $\ket{\phi_1}$ be the corresponding purifications
on the joint system of Alice and Bob. If $\rho_0$ equals $\rho_1$
then Alice can find a local unitary transformation $U$ that she
can apply to her part of the system and satisfying
$\ket{\phi_1} = U \otimes \id \ket{\phi_0}$. This enables Alice to change the total
state from $\ket{\phi_0}$ to $\ket{\phi_1}$ and thus cheat. This also holds in an approximate
sense~\cite{mayers:trouble}, used here in the following
form:
\begin{lemma}\label{lemma:mayers}
Let $\delta(\rho_0, \rho_1)\leq \epsilon$ and assume that the
bit-commitment protocol is error-free if both parties are honest. Then there is a
method for Alice to cheat such that the probability of successfully
revealing a $0$ given that she committed to a $1$ is greater or
equal to $1-\sqrt{2\epsilon}$.
\end{lemma}
\begin{proof}
$\delta(\rho_0, \rho_1) \leq \epsilon$ implies $F(\rho_0, \rho_1)
\geq 1-\epsilon$. $F(\cdot, \cdot)$ is the fidelity of two quantum
states, which equals $\max_{U}| \bra{\phi_0} U
\otimes \id\ket{\phi_1}|$ by Uhlmann's theorem. Here, $\ket{\phi_0}$ and $\ket{\phi_1}$
are the joint states after the commit phase and the maximization
ranges over all unitaries $U$ on Alice's (i.e. the purification)
side. Let $\ket{\psi_0} = U \otimes \id \ket{\phi_1}$ for a $U$
achieving the maximization. Then
\begin{eqnarray*}
\delta(\proj{\phi_0},
\proj{\psi_0})&=&\sqrt{1-|\braket{\phi_0}{\psi_0}|}\\
&\leq& \sqrt{1-(1-\epsilon)^2}\\
&\leq& \sqrt{2\epsilon}.
\end{eqnarray*}
If both parties are
honest, the reveal phase can be regarded as a measurement resulting
in a distribution $P_Y$ ($P_Z$) if $\ket{\phi_0}$ ($\ket{\psi_0}$)
was the state before the reveal phase. The random variables $Y$ and
$Z$ carry the opened bit or the value `reject (r)'. Since the trace
distance does not increase under measurements, $\delta(P_Y, P_Z)
\leq \delta(\proj{\phi_0}, \proj{\psi_0}) \leq \sqrt{2\epsilon}$.
Hence $\frac{1}{2}(|P_Y(0)-P_Z(0)| +
|P_Y(1)-P_Z(1)|+|P_Y(r)-P_Z(r)|)  \leq \sqrt{2\epsilon}$. Since
$\ket{\phi_0}$ corresponds to Alice's honest commitment to $0$ we
have $P_Y(0)=1$, $P_Y(1)=P_Y(r)=0$ and hence $P_Z(0)\geq
1-\sqrt{2\epsilon}$.
\end{proof}

\section{Impossibility}\label{imposs}
The proof of our impossibility result consists of three steps:
in the previous section, we saw that any $(n,a,b)$-QBSC is also
an $(n,a,b)$-$\mbox{QBSC}_\xi$ with the security measure $\xi(\ens)$
defined eq.~(\ref{xi}). Below, we prove that an
$(n,a,b)$-$\mbox{QBSC}_\xi$ can only exist for values $a$, $b$ and
$n$ obeying $a + b + c \geq n$, where $c$ is a small constant
independent of $a$, $b$ and $n$. This in turn implies the
impossibility of an $(n,a,b)$-QBSC for such parameters. At the end of this section we
show that $\emph{many}$ executions of the protocol can only be
secure if $a + b \geq n$.

The intuition behind our main argument is simple: To cheat, Alice first chooses
a two-universal hash function $g$. She then commits to a superposition
of all strings for which $g(x) = y$ for a specific $y$. We know from
the privacy amplification theorem above, however, that even though Bob may gain
some knowledge about $x$, he is entirely ignorant about $y$. But then
Alice can change her mind and move to a different set of strings for which
$g(x) = y'$ with $y \neq y'$ as we saw above! The following figure
illustrates this idea.

\begin{figure}[h]
\begin{center}
\includegraphics[scale=0.6]{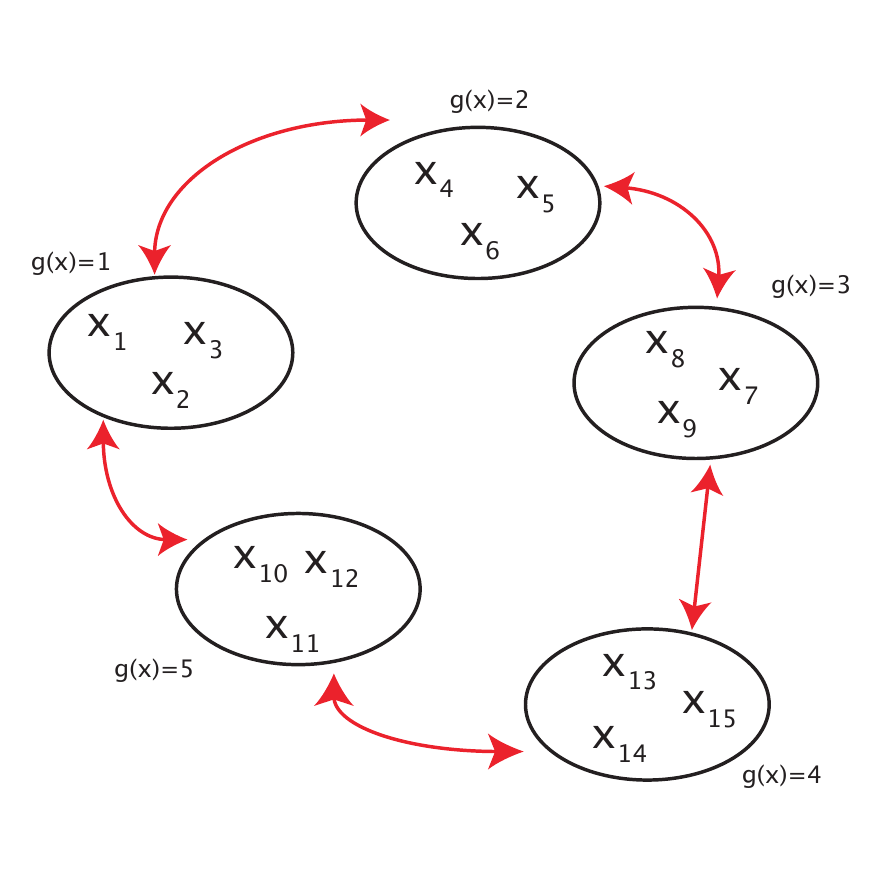}
\caption{Moving from $y$ to $y'$.}
\end{center}
\end{figure}

\begin{theorem} \label{nogo}$(n,a,b)$-$\mbox{QBSC}_\xi$ schemes, and thus
also $(n,a,b)$-$\mbox{QBSC}$ schemes, with $a+b+c < n$ do not exist.
$c$ is a constant equal to $5 \log_2 5 - 4\approx 7.61$.
\end{theorem}
\begin{proof}
Consider an $(n,a,b)$-$\mbox{QBSC}_\xi$ and the case
where both Alice and Bob are honest. Alice committed to
$x$. We denote the joint state of the Alice-Bob-Channel system
$\hilbert_A \otimes \hilbert_B \otimes \hilbert_C$ after the commit
phase by $\ket{\phi_x}$ for input state $\ket{x}$.
Let $\rho_x$ be Bob's reduced density matrix and let
$\cE=\{p_x, \rho_x\}$
where $p_x=2^{-n}$.

Assuming that Bob is honest, we will give a cheating strategy for
Alice in the case where $a+b+5 \log_2 5 - 4 < n$. The strategy will
depend on the two-universal hash function $g:\mathcal{X} = \01^n \rightarrow
\mathcal{Y} = \01^{n -m}$, for appropriately chosen $m$. Alice
picks a $y\in \mathcal{Y}$ and prepares the state $ (\sum_{x \in
g^{-1}(y)}\ket{x}\ket{x})/\sqrt{|g^{-1}(y)|}$. She then gives the
second half of this state as input to the protocol and stays honest
for the rest of the commit phase. The joint state of Alice and Bob
at the end of the commit phase is thus $\ket{\psi^g_y} =
(\sum_{x \in g^{-1}(y)}\ket{x}\ket{\phi_x})/\sqrt{|g^{-1}(y)|}$. The
reduced states on Bob's side are
$\sigma^g_y = \frac{1}{q^g_y}\sum_{x \in g^{-1}(y)} p_x \rho_x$
with probability $q^g_y = \sum_{x \in g^{-1}(y)} p_x$. We
denote this ensemble by $\cE_g$. Let $\sigma =
\sigma^g = \sum_y q^g_y \sigma^g_y$ for all $g$.

We now apply Theorem~\ref{theorem:renato} with $s=n-m$ and
$\xi(\ens)\leq b$ to obtain $ \frac{1}{|\mathcal{G}|} \sum_{g \in
\mathcal{G}} d(\cE_g) \leq \eps $
where $\eps=\frac{1}{2}2^{-\frac{1}{2}(m-b)}$. Hence, there is at
least one $g$ such that $d(\cE_g) \leq \eps$; intuitively, this
means that Bob knows only very little about the value of $g(x)$.
This $g$ defines Alice's cheating strategy. It is straightforward to verify that
$d(\cE_g) \leq \eps$ implies
\begin{equation}\label{sum_eq}
2^{-(n-m)} \sum_{y} \delta(\sigma, \sigma^g_y) \leq 2\eps.
\end{equation}
Let us therefore assume without loss of generality that Alice
chooses $y_0\in \cY$ with $\delta(\sigma,\sigma^g_{y_0}) \leq
2\eps$.

Clearly, the probability to successfully reveal some $x$ in
$g^{-1}(y)$ given $\ket{\psi^g_y}$ is one. Note that Alice learns
$x$, but can't pick it: she committed to a superposition and $x$ is
chosen randomly by measurement. Thus the probability to reveal
$y$ (i.e. to reveal an $x$ such that $y = g(x)$) given $\ket{\psi^g_y}$
successfully is one. Let $\tilde{p}_x$
and $\tilde{q}^g_y$ denote the probabilities to successfully reveal
$x$ and $y$ respectively and $\tilde{p}^g_{x|y}$ be the conditional
probability to successfully reveal $x$, given $y$. We have
$$
\sum_x \tilde{p}_x =\sum_y \tilde{q}^g_y
\sum_{x \in g^{-1}(y)}\tilde{p}^g_{x|y} \geq \sum_y \tilde{q}^g_y.
$$

Recall that Alice can transform $\ket{\psi^g_{y_0}}$ approximately
into $\ket{\psi^g_y}$ if $\sigma^g_{y_0}$ is sufficiently close to
$\sigma^g_y$ by applying local transformations to her part alone. It
follows from Lemma~\ref{lemma:mayers} that we can estimate the
probability of revealing $y$, given that the state was really
$\ket{\psi_{y_0}}$. Since this reasoning applies to all $y$, on
average, we have
\begin{eqnarray*}
\sum_y \tilde{q}^g_y &\geq& \sum_y (1- 2^{\frac{1}{2}} \delta(\sigma^g_{y_0}, \sigma^g_{y})^{\frac{1}{2}})  \\
&\geq& 2^{n-m} - 2^{\frac{1}{2}} 2^{n-m} \big(2^{m-n} \sum_y \delta(\sigma^g_{y_0},\sigma^g_{y})\big)^{\frac{1}{2}} \\
&\geq& 2^{n-m}[1 - 2^{\frac{1}{2}} (2^{m-n}
(\sum_y \delta(\sigma^g_{y_0},\sigma) + \delta(\sigma,\sigma^g_{y})))^{\frac{1}{2}}]\\
&\geq& 2^{n-m}(1 - 2 (2\eps)^{\frac{1}{2}}),
\end{eqnarray*} where the first inequality follows from
Lemma~\ref{lemma:mayers}, the second from Jensen's inequality and
the concavity of the square root function, the third from the
triangle inequality and the fourth from eq.~(\ref{sum_eq}) and
$\delta(\sigma^g_{y_0}, \sigma) \leq 2\eps$. Recall that to be
secure against Alice, we require $2^a \geq 2^{n-m}(1- 2
(2\eps)^{\frac{1}{2}})$. We insert
$\epsilon=\frac{1}{2}2^{-\frac{1}{2}(m-b)}$, define $m=b+\gamma$ and
take the logarithm on both sides to get \be \label{eq-delta} a+b
+\delta \geq n,\ee where $\delta = \gamma -\log_2 (1-2^{-\gamma/4
+1})$. Keeping in mind that $1-2^{-\gamma/4 +1}>0$ (or equivalently
$\gamma>4$), we find that the minimum value of $\delta$ for which
eq.~(\ref{eq-delta}) is satisfied is $\delta=5 \log_2 5 -4$ and
arises from $\gamma = 4 (\log_2 5 - 1)$. Thus,
no $(n,a,b)$-$\mbox{QBSC}_\xi$ with $a+b+5 \log_2 5 - 4 < n$ exists.
\end{proof}

Since the constant $c$ does not depend on $a$, $b$ and $n$, multiple
parallel executions of the protocol in the form of multiple simultaneous commit phases followed by the corresponding opening phases, can only be secure if $a+b\geq
n$:
\begin{proposition} Let $P$ be an $(n,a,b)$-$\mbox{QBSC}_\xi$ or $(n,a,b)$-$\mbox{QBSC}$.
The $m$-fold parallel execution of $P$ will be insecure if $a+b<n-c/m$. In particular, no $(n,a,b)$-$\mbox{QBSC}_\xi$ or $(n, a,
b)$-QBSC with $a+b<n$ can be executed securely an arbitrary number
of times in parallel. Furthermore, no $(n,a,b)$-$\mbox{QBSC}_\chi$ with $a+b<n$ and $\chi$ the Holevo information can be executed securely an arbitrary number of times in parallel.
\end{proposition}
\begin{proof}
In the following, we assume wlog that $a$ and $b$ are the smallest cheat parameters for $P$.
Let $Q$ denote the $(nm, a_m, b_m)$-$\mbox{QBSC}_\xi$ or $(nm, a_m, b_m)$-$\mbox{QBSC}$ protocol obtained by executing $P$ $m$ times in parallel. By Theorem~\ref{nogo}, $Q$ is insecure if $a_m+b_m<nm-c$. Since $a$ and $b$ were assumed to be the smallest cheat parameters for $P$, the product cheating attack by Alice and Bob lead to the estimates $a_m\geq a m$ and $b_m\geq bm$, respectively. Therefore, the $m$ fold execution of $P$ is insecure, if $am+bm\leq a_m+b_m<nm-c$ or $a+b<n-c/m$.

In order to prove the result about Holevo information QBSC, we will use a slightly different characterisation of privacy amplification in the proof of Theorem~\ref{nogo}. In this characterisation, the right hand side of eq.~(\ref{eq-renner-koenig}) is replaced by $\kappa+2^{-\frac{1}{2}[H_{\min}^\kappa(\rho_{AB}|\rho_B)-s]}$ for an arbitrary $\kappa> 0$~\cite[Corollary 5.6.1]{renato:diss}. Going through the proof with this change in mind, one sees that $Q$ is not a $(nm, a_m, b_m)$-$\mbox{QBSC}_{\Xi}$ for $\Xi(\tilde\cE)=nm-H_{\min}^\kappa(\tilde\rho_{AB}|\tilde\rho)$ if $a_m+b_m+\delta\leq mn $. Here, $\tilde \cE$ is the ensemble corresponding to $Q$ and $\tilde\rho_{AB}$ and $\tilde\rho$ the related states; $\delta\equiv \delta(\kappa)$ is a positive constant independent of $n$. Since $\tilde\cE=\cE^{\otimes m}$ and thus $\tilde\rho_{AB}=\rho_{AB}^{\otimes m}$ and $\tilde\rho_{AB}=\rho_{AB}^{\otimes m}$ we are able to invoke the estimate
$$\frac{1}{m}H_{\min}^\kappa(\rho_{AB}^{\otimes m}|\rho^{\otimes m}) \geq H(\rho_{AB})-H(\rho)-3\lambda$$ where $\lambda(\kappa, m) \rightarrow 0$ as $m \rightarrow \infty$~\cite[Chain rule in Theorem 3.1.12 and Theorem 3.3.4]{renato:diss} in order to conclude that $Q$ is not a $(nm, a_m, b_m)$-$\mbox{QBSC}_{m (\chi(\cE)+2\lambda)}$ if $a_m+b_m+\delta< mn $. This shows that if $P$ is a $(nm, a_m, b_m)$-$\mbox{QBSC}_{m (\chi(\cE)+2\lambda)}$ with
$\alpha_m m+\beta_m m\leq a_m+b_m<nm-\delta$, i.e. $\alpha_m +\beta_m <n-\delta/m,$ then its $m$-fold execution cannot be secure. Taking $m$ to infinity we see that if $P$ is an $(n,a,b)$-$\mbox{QBSC}_\chi$ with $a+b<n$ then it cannot be executed securely an arbitrary number of times in parallel.
\end{proof}

It follows directly from~\cite{kitaev:super} that
the results in this section also hold in the presence of
superselection rules.

\section{Possibility}\label{protocol}
Surprisingly, if one is willing to
measure Bob's ability to learn $x$ using the accessible information,
non-trivial protocols become possible.
These protocols are based on
a discovery known as ``locking of classical information in quantum
states''~\cite{terhal:locking}.

\subsection{A Family of Protocols}
The protocol, which we call LOCKCOM($n$, $\cU$), uses this effect
and is specified by a set $\cU = \{U_1,\ldots,U_{|\cU|}\}$ of
unitaries.
\begin{itemize}
\item Commit phase: Alice has the string $x \in \01^n$ and randomly chooses
$r \in \{1,\ldots,|\cU|\}$. She sends the state $U_r \ket{x}$ to
Bob, where $U_r \in \cU$.
\item Reveal phase: Alice announces $r$ and $x$. Bob applies
$U_r^\dagger$ and measures in the computational basis
to obtain $x'$. He accepts if and only if $x'=x$.
\end{itemize}

We first show that our protocol is secure with respect to Definition~\ref{securityDef}
if Alice is dishonest. Note that our proof only depends on the number of unitaries used,
and is independent of a concrete instantiation of the protocol.
\begin{lemma}\label{Alice-lemma}
Any LOCKCOM$(n,\cU)$ protocol is $\log(|\cU|)$-binding, i.e. $2^a \leq
|\cU|$,
\end{lemma}
\begin{proof}
Let $p^A_x$ denote the probability that Alice reveals $x$ successfully. Then,
$p^A_x \leq \sum_r p^A_{x,r}$, where $p^A_{x,r}$
is the probability that $x$ is accepted by Bob when the reveal
information was $r$. Let $\rho$ denote the state of Bob's system.
Summation over $x$ now yields
\begin{eqnarray*}
\sum_x p^A_x &\leq&  \sum_{x, r}p^A_{x,r} \\
&=& \sum_{x,r} \Tr \proj{x} U_r^\dagger \rho U_r \\
&=& \sum_r \Tr \rho = |\cU|,
\end{eqnarray*}
hence $a \leq \log_2 |\cU|$
\end{proof}

In order to examine security against a dishonest Bob, we have to consider the actual form
of the unitaries. We first show that there do indeed exist interesting protocols. Secondly,
we present a simple, implementable, protocol.
To see that interesting protocols can exist, let Alice choose a set of $O(n^4)$ unitaries independently
according to the Haar measure (approximated) and announce the resulting set $\cU$
to Bob. They then perform LOCKCOM($n,\cU$).
Following the work of~\cite{winter:randomizing}, we now show that this variant is secure
against Bob with high probability in the sense that there exist $O(n^4)$ unitaries that bring Bob's
accessible information down to a
constant: $\I(\cE) \leq 4$:
\begin{theorem} \label{theorem-LOCKCOM}\label{randomProtocol}
For $n \geq 3$, there exist $(n, 4\log_2 n +O(1), 4)$-$ \mbox{QBSC}_{\I}$ protocols.
\end{theorem}
\begin{proof}
Let $\cU_{ran}$ denote the set of $m$ randomly chosen bases
and consider the LOCKCOM($n,a,b$) scheme using unitaries $\cU = \cU_{ran}$.
Security against Alice is again given by Lemma~\ref{Alice-lemma}. We now need
to show that this choice of unitaries achieves the desired locking effect and
thus security against Bob.
Again, let $d=2^n$ denote the dimension. It was observed
in \cite{terhal:locking} that
$$
\I\leq \log_2 d +\max_{\ket{\phi}} \sum_i \frac{1}{m} H(X_j),
$$
where $X_j$ denotes the outcome of the measurement of $\ket{\phi}$ in basis
$j$ and the maximum is taken over all pure states $\ket{\phi}$.
According to \cite[Appendix B]{winter:randomizing} there is a
constant $C'>0$ such that
\begin{eqnarray*}
\Pr [\inf_{\ket{\phi}} \frac{1}{m} \sum_{j=1}^m H(X_j) &\leq& (1-\e) \log_2 d -3 ]\\
&\leq& \left( \frac{10}{\epsilon} \right)^{2d}
2^{ -m \left(\frac{\epsilon C' d}{2(\log_2 d)^2} -1\right)
},
\end{eqnarray*}
for $d \geq 7$ and $\epsilon \leq 2/5$. Set $\epsilon=\frac{1}{\log_2 d}$.
The RHS of the above equation then decreases provided that $m > \frac{8}{C'} (\log_2 d)^4$.
Thus with $d=2^n$ and $\log_2 m=4 \log_2 n + O(1)$,
the accessible information is then $\I \leq \log_2 d -
(1-\epsilon) \log_2 d +3 = \epsilon \log_2 d +3=4$ for our choice
of $\epsilon$.
\end{proof}

Unfortunately, the protocol is inefficient both in terms of
computation and communication. It remains open to find an efficient
constructive scheme with those parameters.

In contrast, for only two bases, an efficient
construction exists and uses the identity and the Hadamard transform
as unitaries. For this case, the security of the standard LOCKCOM
protocol follows immediately:

\begin{theorem} \label{two-bases}
LOCKCOM($n,\{\id^{\otimes n}, H^{\otimes n}\}$) is a $(n, 1, n/2)-\mbox{QBSC}_{\I}$
protocol.
\end{theorem}
\begin{proof}
It is sufficient to apply Lemma $\ref{Alice-lemma}$ and the fact that for Bob
$\I \leq n/2$~\cite{terhal:locking,CW05}.
\end{proof}

\section{A cheat-sensitive protocol}\label{CSprotocol}

\subsection{Scenario and Result}

We now extend the protocol above to be cheat-sensitive against Bob.
That is, even though Bob may be able to gain a lot of information on
the committed string, Alice has a decent probability of catching Bob
if he actually tries to extract such information~\footnote{The
results in this section are included in~\cite{matthias:thesis}.}.

We first extend our definition to accommodate cheat-sensitivity
against Bob.

\begin{definition}
A $(n,a,b)$-$B$-QBSC is \emph{cheat-sensitive} against Bob if there
is a non-zero probability that he will be detected by Alice when he
cheats.
\end{definition}

We elaborate below on the scenario in which we analyse Bob's
cheating and thus make precise what we mean by saying \emph{Bob
cheats}.

The following protocol is a modification of LOCKCOM($n, {\cal U}$)
which incorporates cheat-sensitivity against Bob.

\begin{protocol}{CS-Bob-LOCKCOM($n,
{\cal U}$)}{}\index{LOCKCOM}\label{lockcom}
\item Commit phase: Alice randomly chooses the string
$x \in \01^n$ and a unitary $U_r$ from a set of unitaries ${\cal U}$
known to both Alice and Bob. She sends the state $U_r \ket{x}$.
\item Reveal phase: Alice sends $r$ to Bob, he applies
$(U_r)^\dagger$ to the state that he received from Alice and
measures in the computational basis. His outcome is denoted by $y$.
\item Confirmation phase: Bob sends $y$ to Alice. If Alice is honest, and if
$x=y$ she declares `accept' otherwise `abort'.
\end{protocol}

We proved in Theorem~\ref{two-bases} that CS-Bob-LOCKCOM($n,
\{\id^{\otimes n}, H^{\otimes n}\}$) is a $(n, 1, n/2)$-$\I$-quantum
string commitment protocol. In fact this result can be extended to
dimensions different from $d=2^n$ where one can show that
CS-Bob-LOCKCOM($\log_2 d, \{\id, U\}$), where $U$ is the Fourier
transform, is a $(\log_2 d, 1, \frac{\log_2 d}{2})$-$\I$-quantum
string commitment protocol.

We now restrict our attention to this protocols and prove that a
dishonest Bob is detected whenever he has obtained a non-zero amount
of information about $x$ \emph{before} the reveal
stage~\footnoteremember{myfoot}{This information is stored in a
register $C$ and not touched upon later on; Bob's remaining
information is called $Q$. Note that this cheating scenario includes
cheating by measurement, since here $C$ contains the classical
measurement result of which he can put a copy into $Q$. Any later
manipulation of $Q$ can therefore be achieved without touching
$C$.}. More precisely, we give a tradeoff for cheat detection versus
Holevo-information gain against a dishonest Bob, with the property
that every nonzero Holevo-information gain leads to a nonzero
detection probability of Bob.

\begin{theorem} \label{theorem-Bob-cheat-sensitive}
If Bob is detected cheating with probability less than $\epsilon$,
then his \emph{Holevo information gain} obeys
$$\chi(\cE^C) \leq
4\sqrt{\epsilon}\log_2 d +2\mu(2\sqrt{\epsilon}).$$
\end{theorem}

As a corollary we find that CS-Bob-LOCKCOM($\log_2 d, \{\id, U\}$)
is cheat sensitive against Bob.

\begin{corollary}
Bob will be detected cheating with a nonzero probability, if he
gathers a nonzero amount of Holevo information.
\end{corollary}

\subsection{Proof}
\begin{figure}
\begin{center}
\vspace{0cm}
\includegraphics[width=0.4\textwidth]{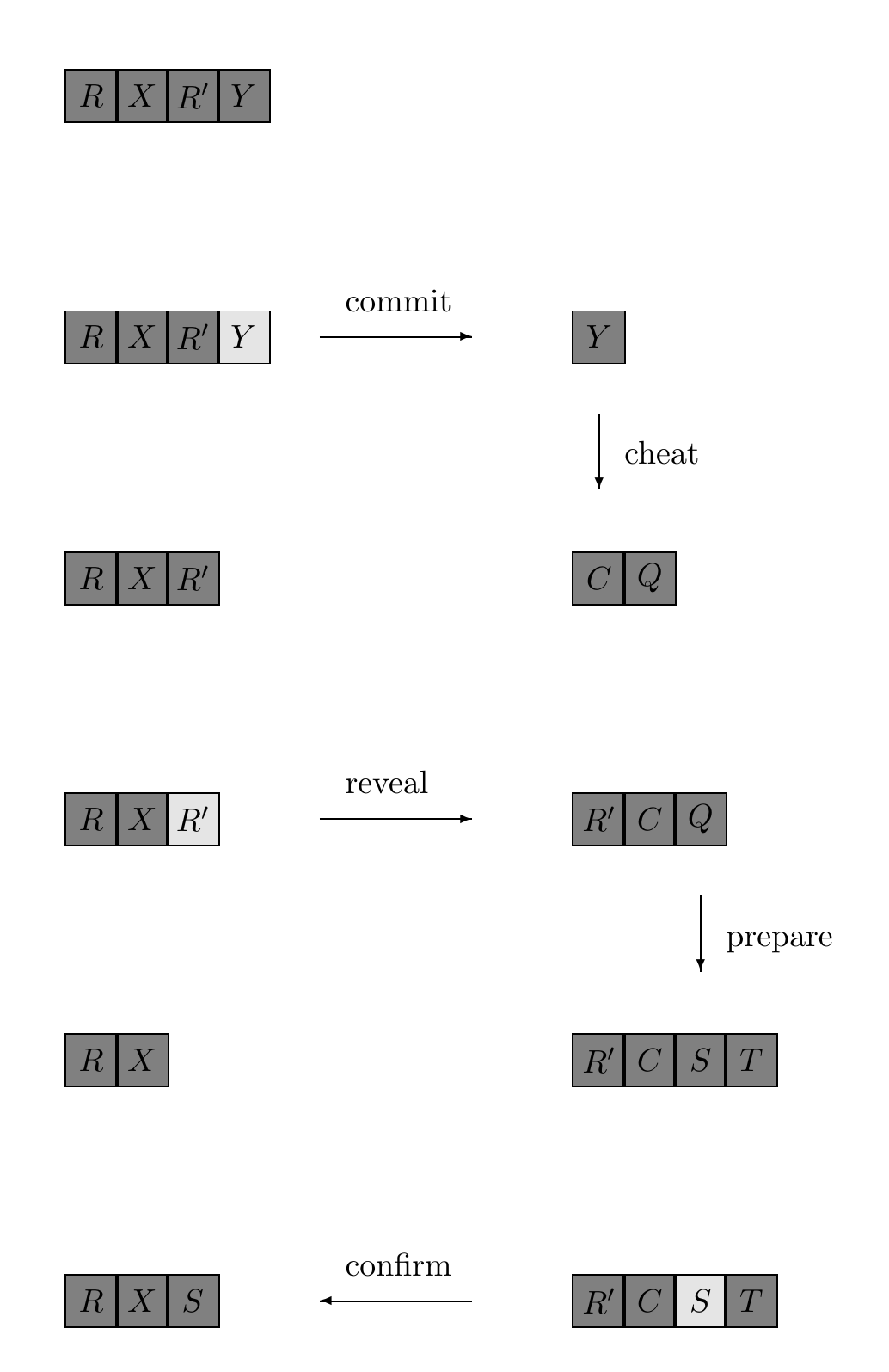}
\end{center}
\caption{Execution of CS-Bob-LOCKCOM with honest Alice on the left and
cheating Bob on the right. Time flows downwards.}
\label{figure-cheat}
\end{figure}

We start this section with a description of the sequence of events
for the case where Alice is honest and Bob applies a general
cheating strategy (see also Figure~\ref{figure-cheat}).

\begin{itemize}
\item The commit phase of the protocol $LOCKCOM(\log_2 d, \{\id, U\})$ is equivalent to the following procedure:
Alice prepares the state
$$\ket{\psi}=\frac{1}{\sqrt{2d}} \sum_{x,r} \ket{x}^X\ket{r}^R\ket{r}^{R'}U^r
\ket{x}^Y$$ on the system $XRYR'$ and sends system $Y$ (over a
noiseless quantum channel) to Bob. It is understood that $U^0=\id$
and $U^1=U$. Note that $R'$ contains an identical copy of $R$ and
corresponds to the \emph{reveal information}.
\item Bob's most general cheating operation can be described by a unitary matrix $V_{cheat}$ that splits the system $Y$
into $C$ and $Q$. $C$ contains by definition the information
gathered during cheating and is not touched upon later on
\footnoterecall{myfoot}.
$$V_{cheat}: Y \rightarrow CQ$$
The map $V_{cheat}$ followed by the partial trace over $Q$ is
denoted by $\Lambda^C$ and likewise $V_{cheat}$ followed by the
partial trace over $C$ is denoted by $\Lambda^Q$.
\item Alice sends the reveal information $R'$ to Bob.
\item Bob applies a preparation unitary $V_{prepare}$ to his
system. Since $C$ will not be touched upon, the most general
operation acts on $R'Q$ only:
$$V_{prepare}: R'Q \rightarrow R'ST.$$
Bob then sends $S$ to Alice and keeps $T$.
\item Alice measures $S$ in the computational basis and compares
the outcome to her value in $X$. If the values do not agree, we say
that \emph{Alice has detected Bob cheating}. The probability for
this happening is given by
$$ \frac{1}{d}\sum_{x=1}^d \left(1-\Tr \proj{x} \rho^S_x\right),$$
where $\rho^S_x=\Tr_{XRR'T} \proj{x} \proj{\psi}^{XRR'ST}$, and
$\ket{\psi}^{XRR'ST}$ is the pure state of the total system after
Bob's application of $V_{prepare}$.
\end{itemize}

\noindent Note that Alice measures in the computational basis since
for honest Bob $V_{prepare}=\sum_{r' \in \01} \proj{r'} \otimes
(U^r)^\dagger$, in which case his outcome agrees with the committed
value of an honest Alice.

Before we start with the proof of
Theorem~\ref{theorem-Bob-cheat-sensitive}, we define ensembles
depending on the classical information contained in $XR$, i.e. for
$Z \in \{C, Q \}$, define $\cE^Z_{r}=\{p_{x}, \rho_{xr}^Z\}$  with
$$
\rho_{xr}^Z=\frac{1}{p_x p_r}\Tr_{XRR'CQ\backslash Z} \proj{xr} \proj{\psi}^{XRR'CQ}
$$
and for $Z \in \{S, T \}$ let $\cE^Z_{r}=\{p_{x}, \rho_{xr}^Z\}$ with
$$
\rho_{xr}^Z=\Tr_{XRR'CST\backslash Z} \proj{xr} \proj{\psi}^{XRR'CST}.
$$
Sometimes we are only interested in the ensemble averaged over the
values of $r$: for $ Z \in \{C, Q, S, T\}$
    \be
        \cE^Z=\{p_{x},  \rho_x^Z\} \textrm{ where } \rho_x^Z= \half \left(\rho_{x0}^Z+
        \rho_{x1}^Z \right).
    \ee

Let us now come to two technical lemmas, most notably a channel
uncertainty relation (Lemma~\ref{lemma-uncertainty}) that was
discovered in connection with squashed entanglement:

Consider a uniform ensemble $\cE_0=\{ \frac{1}{d}, \ket{i}
\}_{i=1}^{d}$ of basis states of a Hilbert space $\cH$ and the
ensemble $\cE_1=\{ \frac{1}{d}, U\ket{i} \}_{i=1}^{d}$ rotated with
a unitary $U$. Application of the completely positive trace
preserving (CPTP) map $\Lambda$ (with output in a potentially
different Hilbert space) results in the two ensembles
\begin{align*}
  \Lambda(\cE_0) &= \left\{ \frac{1}{d}, \Lambda(\proj{i}) \right\}         \\
  \Lambda(\cE_1) &= \left\{ \frac{1}{d}, \Lambda(U\proj{i}U^\dagger) \right\}
\end{align*}
with Holevo information for $\cE_0$ given by
$$\chi(\Lambda(\cE_0)) = H\left( \frac{1}{d}\sum_i \Lambda(\proj{i}) \right)
                         - \frac{1}{d}\sum_i H\bigl( \Lambda(\proj{i}) \bigr)$$
and similarly for $\cE_1$. Consider also the quantum mutual
information of $\Lambda$ relative to the maximally mixed state $\tau
= \frac{1}{d}\id$, which is the average state of either $\cE_0$ or
$\cE_1$:
$$I(\tau;\Lambda) = H\bigl(\tau\bigr)
                   + H\bigl(\Lambda(\tau)\bigr)
                   - H\bigl((\id\otimes\Lambda) (\proj{\psi_d})\bigr),$$
where $\ket{\psi_d}$ is a maximally entangled state in dimension $d$
purifying $\tau$.

\begin{lemma}[Channel Uncertainty Relation\index{channel uncertainty relation}~\cite{CW05}]
  \label{lemma-uncertainty}
  Let $U$ be the Fourier transform of dimension $d$, i.e. of
  the Abelian group $\integers_d$ of integers modulo $d$. More
  generally,
  $U$ can be a Fourier transform of any finite Abelian group labeling
  the ensemble $\cE_0$, e.g. for $d=2^\ell$, and the group
  $\integers_2^\ell$, $U=H^{\otimes\ell}$
  with the Hadamard transform $H$ of a qubit.
  Then for all CPTP maps $\Lambda$,
  \be
    \label{channel-ineq}
    \chi\bigl(\Lambda(\cE_0)\bigr) + \chi\bigl(\Lambda(\cE_1)\bigr)
                                                   \leq I(\tau; \Lambda).
  \ee
\end{lemma}

The following technical lemma is a technical consequence of Fannes'
inequality.

\begin{lemma} \label{lemma-Holevo-relation}
Let $\cE=\{p_i, \rho_i=\proj{\psi_i}\}$ be an ensemble of pure
states and $\tilde{\cE}=\{p_i, \sigma_i\}$ be an ensemble of mixed
states, both on $\complex^d$. If $\sum_i p_i \bra{\psi_i} \sigma_i
\ket{\psi_i} \geq 1-\epsilon$, then
$$ |\chi(\tilde{\cE}) - \chi(\cE)| \leq 4\sqrt{\epsilon} \log_2 d  +2 \mu(2\sqrt{\epsilon}),$$
where $\mu(x)=\min \{-x\log_2 x, \frac{1}{e}\}$.
\end{lemma}

\begin{proof}
The justification of the estimate
$$ \epsilon \geq \sum_i p_i (1-\Tr \rho_i \sigma_i) \geq \sum_i
p_i \delta_i^2 \geq \big(\sum_i p_i \delta_i \big)^2,$$ where
$\delta_i= \delta(\rho_i, \sigma_i)$ is as follows: the second
inequality is a standard relation between the fidelity and the trace
distance and the third follows from the convexity of the square
function. Strong convexity of the trace distance implies
$\delta(\rho, \sigma) \leq \sqrt{\epsilon}$. Fannes' inequality will
be applied to the overall state
$$ |H(\rho)-H(\sigma)| \leq 2\sqrt{\epsilon} \log_2 d +  \min \{\eta(2 \sqrt{\epsilon}), \frac{1}{e}\}$$
where $\eta(x)=-x\log_2 x$, and to the individual ones \beastar
\sum_i p_i |H(\sigma_i)- H(\rho_i)| &\leq& \big(\sum_i p_i \delta_i
\big) 2\log_2 d  +\\
&&\sum_i p_i \min
\{\eta(2 \delta_i), \frac{1}{e}\}\\
&\leq & \sqrt{\epsilon} 2\log_2 d + \min \{\eta(2 \sqrt{\epsilon}),
\frac{1}{e}\} \eeastar where the last inequality is true by the
concavity of $\eta(x)$. Inserting these estimates in the Holevo
$\chi$ quantities $\chi(\cE)=H(\rho)$ and
$\chi(\tilde{\cE})=H(\sigma)-\sum_i p_i H(\sigma_i)$ concludes the
proof.
\end{proof}

\begin{proof}[Proof of Theorem~\ref{theorem-Bob-cheat-sensitive}]
Let $\cE_0$ and $\cE_1$ be defined as in
Lemma~\ref{lemma-uncertainty}. In the commit phase of the protocol,
Alice chooses one of the ensembles (each with probability $\half$),
and one of the states in the ensemble (each with probability
$\frac{1}{d}$). The justifications for the following estimate are
given in a list below. \bea
&&\chi(\cE^C_{0})+\chi(\cE^C_{1})\\
 \label{eq-rewriting1}
  &=&\chi(\Lambda^C(\cE_{0}))+\chi(\Lambda^C(\cE_{1}))\\
 \label{eq-rewriting2}&\leq& I(XRR';C) \\
 \label{eq-rewriting3}&=&2H(XRR')-I(XRR';Q)\\
 \label{eq-rewriting4}&\leq& 2H(XRR')-\chi(\Lambda^Q(\cE_{0}))-\chi(\Lambda^Q(\cE_{1}))\\
 \label{eq-rewriting5}&=& 2H(XR)-\chi(\cE^Q_{0})-\chi(\cE^Q_{1})\\
 \label{eq-rewriting6}&\leq& 2H(XR)-\chi(\Lambda^S_0(\cE^Q_{0}))-\chi(\Lambda^S_1(\cE^Q_{1}))\\
 \label{eq-rewriting7}&=& 2H(XR)-\chi(\cE^S_{0})-\chi(\cE^S_{1})\\
 \label{eq-rewriting8}&\leq&2H(XR)-2\chi(\cE^S).
\eea The justifications:
\begin{itemize}
\item Equality~(\ref{eq-rewriting1}): By definition of the string commitment scheme and the map $\Lambda^C$: $\cE^C_r=\{ p_x, \rho_{xr}^C\}=\{ p_x,
\Lambda^C( U^r\proj{x}(U^\dagger)^r)\}=: \Lambda^C(\cE_r)$.
\item Inequality~(\ref{eq-rewriting2}): Application of Lemma~\ref{lemma-uncertainty} for the map $\Lambda^C$. Note that system
$XRR'$ is a reference system for the completely mixed state on
system $Y$ on which the channel $\Lambda^C$ is applied. Hence
$I(\tau; \Lambda^C)=I(XRR';C)$.
\item Equality~(\ref{eq-rewriting3}): Simple rewriting of the entropy
terms making use of the definition of quantum mutual information and
the purity of $XRR'CQ$.
\item Inequality~(\ref{eq-rewriting4}): Application of Lemma~\ref{lemma-uncertainty} for the map $\Lambda^Q$. Note that system
$XRR'$ is a reference system for the completely mixed state on
system $Y$ on which the channel $\Lambda^Q$ is applied. Hence
$I(\tau; \Lambda^Q)=I(XRR';Q)$.
\item Equality~(\ref{eq-rewriting5}): $R'$ is a copy of $R$: $H(XRR')=H(XR)$. By definition of the string commitment scheme and the map $\Lambda^Q$: $\cE^Q_r=\{ p_x,
\rho_{xr}^Q\}$ \linebreak $=\{ p_x, \Lambda^Q(
U^r\proj{x}(U^\dagger)^r)\}.$
\item Inequality~(\ref{eq-rewriting6}) and equality~(\ref{eq-rewriting7}): follow from the data processing inequality
$\chi(\Lambda^H(\cE^Q_r)) \leq \chi(\cE^Q_r)$ and from the
definition $\Lambda^H(\cE^Q_r)=\cE^S_r$.
\item Inequality~(\ref{eq-rewriting8}): Finally $\cE^S=\{p_x,
\rho^S_x=\half \left(\rho^S_{x0}+\rho^S_{x1} \right)\}$, which by
the concavity of von Neumann entropy implies $\chi(\cE^S) \leq \half
\left(\chi(\cE^S_0)+\chi(\cE^S_1)\right)$.
\end{itemize}
If Bob is detected cheating with probability less than $\epsilon$,
then by Lemma~\ref{lemma-Holevo-relation} the Holevo quantity
$\chi(\cE^S)$ of the ensemble given in $S$ that Bob sends to Alice
obeys
    \be \label{eq-0}
        \chi(\cE^S) \geq (1-4 \sqrt{\epsilon})\log
        d-2\mu(2\sqrt{\epsilon}).
    \ee
Inserting inequality~(\ref{eq-0}) into
inequality~(\ref{eq-rewriting8}) and noting that $H(XR)=H(Y)=\log_2 d$
proves the claim.
\end{proof}

\noindent This proves cheat-sensitivity against Bob for the simplest
protocol of the LOCKCOM family.

\section{Conclusion}
We have introduced a framework for
quantum commitments to a string of bits. Even though
string commitments are weaker than bit commitments, we showed that under strong
security requirements, there are no such non-trivial protocols. A property of
quantum states known as \emph{locking}, however, allowed us to
propose meaningful protocols for a weaker security demand.
Since the completion of our original work~\cite{BCHLW05}, Tsurumaru~\cite{tsu:sc2}
has also proposed a different QBSC protocol within our framework.

Furthermore,
we have shown that one such protocol can be made cheat-sensitive.
It is an interesting open question to derive a tradeoff between
Bob's ability to gain information and Alice's ability to detect him cheating
for the protocol of Theorem~\ref{randomProtocol} as well.

A drawback of weakening the security requirement is that LOCKCOM
protocols are not necessarily composable. Thus, if LOCKCOM is
used as a sub-protocol in a larger protocol, the security of the
resulting scheme has to be evaluated on a case by case basis.
However, LOCKCOM protocols are secure when
executed in parallel. This is a consequence of the definition of
Alice's security parameter and the additivity of the accessible
information~\cite{H73,barbara:hiding1}, and sufficient for many cryptographic
purposes.

However, two important open questions remain: First, how can we construct
efficient protocols using more than two bases? It may be tempting to
conclude that we could simply use a larger number of mutually unbiased bases,
such as given by the identity and Hadamard transform. Yet, it has been shown~\cite{wehner06c}
that using more mutually unbiased bases does not necessarily lead to a better
locking effect and thus better string commitment protocols. Second,
are there any real-life applications for this weak quantum string commitment?

\acknowledgments
We thank J.~Barrett, A.~Broadbent, I.~Damg{\aa}rd,
A.~Kent, S.~Massar, R.~Renner, R.~Spekkens and R.~de Wolf for discussions.
We also thank R.~Jain for discussion on his
work~\cite{Jain05}, where, following our preprint~\cite{BCHLW05}, he
used a different method to prove that $(n,a,b)$-$\mbox{QBSC}_\chi$s
with $a + 16b + 31 < n$ do not exist.

We thank a DAAD Doktorandenstipen\-dium, the EPSRC,
the Magdalene College Cambridge,
CFI, CIFAR, CIPI, CRC,
NSERC, PREA and OIT, the NWO vici
project 2004-2009, EU project RESQ
IST-2001-37559, QAP IST 015848, the FP6-FET Integrated Project SCALA, CT-015714,
the Sloan Foundation and QuantumWorks.

\end{document}